\documentclass[12pt]{iopart}
\usepackage{iopams,epsf,graphicx}  
\def\msun{$M_{\odot}$}

\begin{document}

\title[Characterization of a time-frequency method to detect EMRIs]{Detecting extreme mass ratio inspirals with LISA using time-frequency methods II: search characterization}

\author{Jonathan Gair\dag \footnote[3]{email address: jgair@ast.cam.ac.uk}, Linqing Wen\ddag}
\address{\dag Institute of Astronomy, University of Cambridge, Madingley Road, Cambridge, CB3 0HA, UK}
\address{\ddag Max Planck Institut fuer Gravitationsphysik, Albert-Einstein-Institut Am Muehlenberg 1,  D-14476 Golm, Germany}
\begin{center}
\begin{abstract}
The inspirals of stellar-mass compact objects into supermassive black holes constitute some of the most important sources for LISA.  Detection of these sources using fully coherent matched filtering is computationally intractable, so alternative approaches are required. In a previous paper \cite{lisa5}, we outlined a detection method based on looking for excess power in a time-frequency spectrogram of the LISA data. The performance of the algorithm was assessed using a single `typical' trial waveform and approximations to the noise statistics. In this paper we present results of Monte Carlo simulations of the search noise statistics and examine its performance in detecting a wider range of trial waveforms. We show that typical extreme mass ratio inspirals (EMRIs) can be detected at distances of up to 1--3 Gpc, depending on the source parameters. We also discuss some remaining issues with the technique and possible ways in which the algorithm can be improved.
\end{abstract}
\end{center}



\maketitle

\section{Introduction}
\label{intro}
Astronomical observations indicate that many galaxies host a supermassive black hole (SMBH) in their centre. The inspirals of stellar-mass compact objects into such SMBHs with mass $M\sim\mbox{few}\times10^{5}M_{\odot}$--$10^{7}M_{\odot}$ constitute one of the most important gravitational wave (GW) sources for the planned space-based GW observatory LISA. The large parameter space of possible signals and their long duration makes detection using fully coherent matched filtering computationally intractable, so alternative search techniques are required. One option is to use a semi-coherent matched filtering algorithm and a preliminary analysis of that approach \cite{jon04} suggests that the LISA EMRI detection rate will be dominated by inspirals of $\sim10$ \msun\ BHs onto $\sim10^6$\msun\ SMBHs, and could be as high as $\sim 1000$ in 3--5 years within $\sim 3.5$ Gpc.

Matched filtering algorithms are computationally very expensive and it is therefore valuable to scope out alternatives. In a previous paper \cite{lisa5}, we described a strategy to detect GWs from EMRIs by accumulating the signal power in the time-frequency (t-f) domain. The t-f power spectrum is produced by dividing the data into segments of equal duration and carrying out a Fast Fourier Transform (FFT) on each. The length of the data segments affects the resolution of the spectrogram and must be chosen to allow the evolution of the source to be followed in the t-f plane. This is a search parameter that should eventually be optimized, probably in a source specific way. For now we choose $\sim2$ week segments, as these provide appropriate resolution for typical EMRIs. The choice of segment duration does not significantly affect the computational requirements of the search, in contrast to the semi-coherent approach \cite{jon04} which is limited to $2-3$ week segments by computational cost.

The power spectrum is defined for each segment $i$ and frequency bin $k$ as,
\begin{equation} \fl
P(i,k)= \sum_{\alpha = I,II} \left[\frac{2|(h^{\alpha \, i}_k+n^{\alpha \, i}_k)|^2}{\sigma^2_{n^{\alpha}_k}} \right]
=\sum_{\alpha = I,II} \left[\frac{2 (h^{\alpha \, i}_k)^{2}}{\sigma^2_{n^{\alpha}_k}}+4\frac{\mbox{Re}[h^{\alpha \, i}_k (n^{\alpha \, i}_k)^{*}]} {\sigma^2_{n^{\alpha}_k}} +\frac{2 (n^{\alpha \, i}_k)^{2}}{\sigma^2_{n^{\alpha}_k}} \right],
\end{equation}
\label{power}
where $h_k$ denotes the Fourier amplitude of the signal, $\sigma^2_{n_k}=0.5S_{h}(f)/ (\rmd t^2\rmd f) $ is the expected variance of the noise component $n_k$ at frequency bin $k$, characterized by $S_h(f)$, the strain spectral density of the noise, $\rmd t$ is the sampling cadence and $\rmd f$ is the Fourier bin width. We assume that the LISA noise components, $n_k$, can be approximated by a stationary and Gaussian distribution, with spectral density $S_h(f)$. In addition to the instrumental noise, the LISA data stream will contain an unresolvable astrophysical foreground from galactic white dwarf (WD) binaries. We include the white dwarfs as a noise contribution in $S_h$ in the usual way (\cite{leor04}, equations (52) and (54)). The instrumental noise spectral density is also taken from \cite{leor04}, equation (48). Including the WD foreground in this way assumes it is isotropic on the sky, which is not a good assumption in this case. For the same reason, the WD foreground will not be a stationary noise source on long timescales, but this is probably a reasonable approximation for few week long data segments. In the future, our analysis will be repeated using a more accurate realization of the WD distribution (e.g., \cite{nelemans01}). Generally speaking, EMRIs are not detectable when their frequencies overlap with the WD confusion background, so the results in this paper should not change significantly. However, the location in frequency of the confusion limit will be important, i.e., which WD binaries are resolvable and which are not \cite{timpano05}.

In the low-frequency regime, LISA can be regarded as consisting of two independent Michelson detectors which are denoted I and II \cite{curt98}. We combine the responses from these two detectors to produce the power spectrum. In this first analysis, we add the spectrograms from the I and II responses before performing the search. An alternative algorithm might involve searching each spectrogram separately and looking for coincidences. Such refinements will be investigated in the future. For the combined spectrogram, we calculate the power ``density'', $\rho(i,k)$, by computing the average power within a rectangular box centered at each point ($i,k$), 
\begin{equation}
\rho(i,k)= \frac{1}{m}\,\sum^{n/2}_{a=-n/2} \sum^{l/2}_{b=-l/2} P(i+a,k+b),
\end{equation}
where $n$, $l$ are the lengths of the box edges in the time and frequency dimensions respectively and $m=n\times l$ is the number of data points enclosed. To search for a possible signal, we vary the size of the box, and for each choice of $n$ and $l$, $(n_j, l_j)$ say, we search for any points at which $\rho(i,k)$ exceeds a threshold, $\rho_j$. A detection occurs when there is at least one point with $\rho(i,k) > \rho_j$ for at least one box size. In the first algorithm, we use only box sizes $(n_j, l_j) = (2^{n_t}, 2^{n_f})$, for all possible integer choices of $n_t$ and $n_f$. The thresholds, $\rho_j$, are set to ensure a small overall false alarm probability (FAP) for the search. This will be discussed in more detail later.

The optimal box size for a given source will be large enough to include much of the signal power but small enough to avoid too much noise contribution. The optimum will be source specific due to the wide variation in EMRI waveforms. The inspiral of a $0.6$\msun white dwarf occurs much more slowly than that of a $10$\msun black hole, so in the first case, the optimal box size is likely to be longer in the time dimension. GWs from an inspiral into a rapidly spinning black hole or from a highly eccentric inspiral orbit are characterized by many frequency harmonics, often close together. In that case, a box that is wider in the frequency dimension may perform well. In designing a search, a balance must therefore be struck between having sensitivity to a range of sources and increasing the reach of the search for a specific source.

In section~\ref{noisestats} we discuss the statistics of this search in the absence of a signal. In section~\ref{threshchoice} we discuss how to assign thresholds for the different box sizes and in section~\ref{algperf} we illustrate the performance of the algorithm on a set of trial waveforms. We discuss some remaining issues and possible ways to develop time-frequency techniques in section~\ref{conc}.

\section{Statistics of the search in the absence of a signal}
\label{noisestats}
\subsection{Distribution of maximum power for one box size}
If no signal is present, then the power $P(i,k)$ at each point in the unbinned plane is distributed as a $\chi^2$ with 4 degrees of freedom. There are four degrees of freedom since there are two independent Michelson responses and the real and imaginary parts of $n^i_k$ are independent for each. Over many noise realizations, the total binned power, $m\rho(i,k)$, at a specific point in the spectrogram will be distributed like a $\chi^2$ with $4\,m$ degrees of freedom. The probability that $m\rho(i,k)$ exceeds a threshold $\rho_j$ is therefore $1-Q_{\chi^2_{4m}}(\rho_j)$ (where $Q_{\chi^2_{X}}$ is the cumulative distribution function for a $\chi^2$ with $X$ degrees of freedom). If $q$ independent samples are drawn from such a distribution, the probability that the maximum, $\rho_m$, is less than a specified value, $\rho_{0}$, is $P(\rho_m < \rho_0) = [Q_{\chi^2_{4m}}(\rho_0)]^q$. However, in the binned t-f plane, $m\rho(i,k)$ is not independent at nearby points since neighbouring bins overlap. There are at least $N/m$ independent samples in the plane, but less than $N$ (where $N$ is the total number of pixels in the unbinned spectrogram). In \cite{lisa5}, we used the estimate that there are $N/(m/4)$ independent points. In practice, the distribution of the maximum is not analytically tractable, but it can be computed numerically using Monte Carlo simulations.

\begin{figure}
\centerline{\includegraphics[keepaspectratio=true,width=5in,angle=-90]{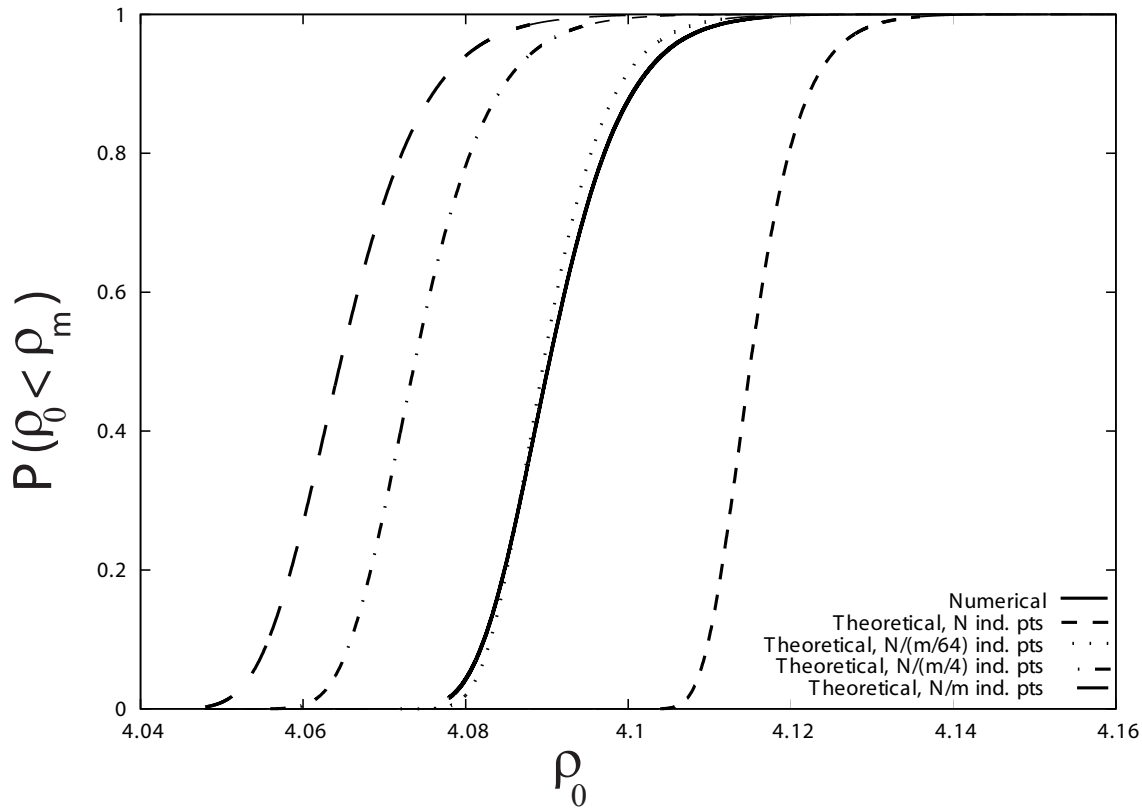}}
\caption{Distribution of maximum power for box size $n = 8$, $l=2048$, as computed from Monte Carlo simulations. Also shown is the theoretical distribution described in the text, under the assumption that there are $N$, $N/(m/64)$, $N/(m/4)$ or $N/m$ independent points in the binned plane.}
\label{sampledist}
\end{figure}

From a Monte Carlo simulation of a quarter of a million noise realizations, we have computed the distribution of the maximum power in the absence of a signal for each box size used. These distributions are much more accurate than the approximations used in \cite{lisa5}. The distributions were computed assuming we had $3 \times 2^{25}$s ($\sim3$ years) of LISA data, sampled at a cadence of $8$s and divided this into $2^{20}$s ($\sim2$ week) sections. This gives a time-frequency spectrogram with $96$ points in time and $65536$ points in frequency and hence $7\times17=119$ possible box sizes of the form $n=2^{n_t}$, $l=2^{n_f}$. For each noise realization, we computed the time-frequency spectrogram of the LISA mission and then calculated the maximum power in each binned spectrogram. Figure~\ref{sampledist} illustrates one typical distribution, with $n=8$ and $l=2048$. It also shows the approximate theoretical distribution described above, under the assumption that there are $q=N$, $N/m$ or $N/(m/4)$ independent samples in the t-f plane. As expected, the true distribution lies between the theoretical distributions for $N$ and $N/m$ independent points. Our first approximation of $N/(m/4)$ independent points \cite{lisa5} is not particularly accurate. Assuming $N/(m/64)$ independent points gives a better approximation to the distribution in this case, although the shape is clearly different. This number is surprisingly large, indicating that boxes with significant overlap are still effectively independent. This suggests that the maximum is usually determined by a few unusually bright pixels, rather than sizable regions of high power. The effective number of independent boxes is box size specific, but always lies between $N/m$ and $N$.

\subsection{Overall search false alarm probability}
\begin{figure}
\centerline{\includegraphics[keepaspectratio=true,height=6in,angle=-90]{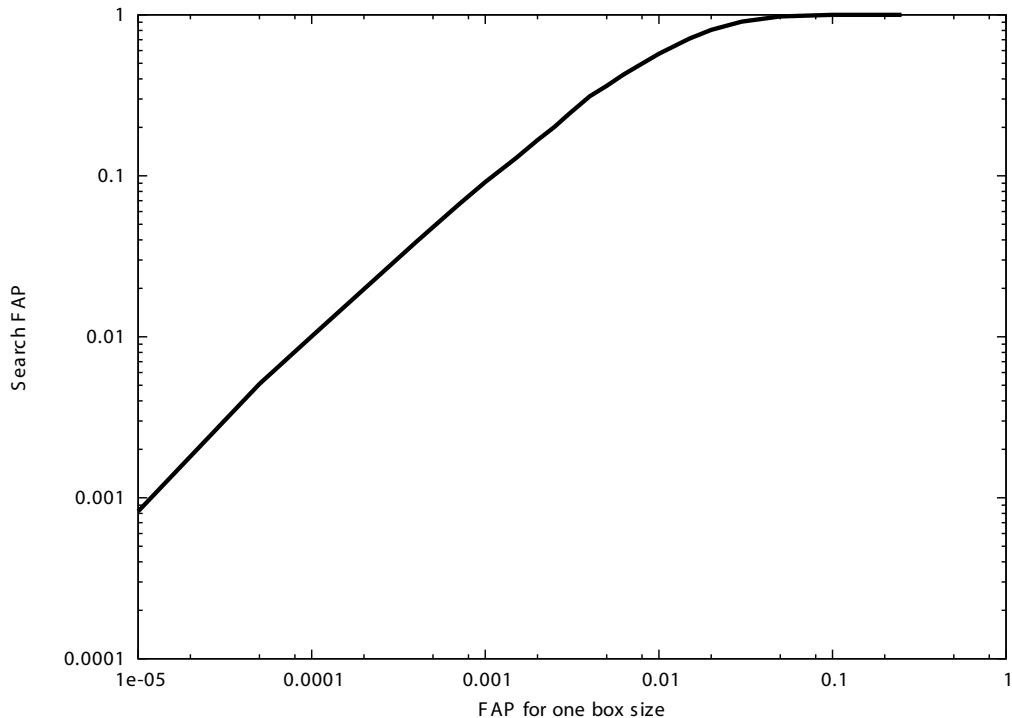}}
\caption{Overall FAP of the search as a function of the FAP assigned to each box size. This assumes all box sizes are assigned equal FAP$_j$.}
\label{searchFAP}
\end{figure}

Assigning a threshold, $\rho_j$, for a given box size amounts to fixing the false alarm probability for that box size, FAP$_j$ (i.e., the probability that this threshold is exceeded by pure noise). If the binned distributions were independent, the overall false alarm probability for the search, FAP, would be FAP = 1 - $\Pi_{j} ( 1-{\rm FAP}_j)$. But as this is not the case, we must make use of the Monte Carlo simulation results again. From the numerical distributions we can compute a threshold for each box size that sets the FAP$_j$ to a specified value. We then use the Monte Carlo data and count the fraction of pure noise realizations in which the maximum power exceeds the threshold for at least one box size. This is the overall search FAP. Since the t-f plane encompasses the whole of the LISA data stream, the FAP is the fraction of LISA missions in which we would expect to have a false alarm. In Figure~\ref{searchFAP}, the search FAP is shown as a function of the (equal) FAP$_j$ assigned to each box size. For small FAP$_j$, FAP$\sim100$FAP$_j$, indicating that effectively $100$ of the $119$ box sizes used are independent.

\section{Choice of thresholds}
\label{threshchoice}
At a fixed search FAP, a threshold choice corresponds to dividing up the FAP among the different box sizes. If we have no prior knowledge of the signal, the natural way to do this is to treat all the box sizes equally, i.e., set FAP$_j=$ constant. This can be done accurately by determining the thresholds from the numerical distributions, as in the previous section. In \cite{lisa5} we set thresholds to give uniform FAP$_j$'s under the theoretical approximation that $\rho^j_m$ was distributed as the maximum of $N/(m_j/4)$ independent samples from a $\chi^2_{4\,m_j}$ distribution. Using that approach, the true FAP$_j$'s are not equal, but we determined the true FAP of the search for each threshold choice using a small Monte Carlo simulation.

\begin{figure}
\centerline{\includegraphics[keepaspectratio=true,width=5in,angle=-90]{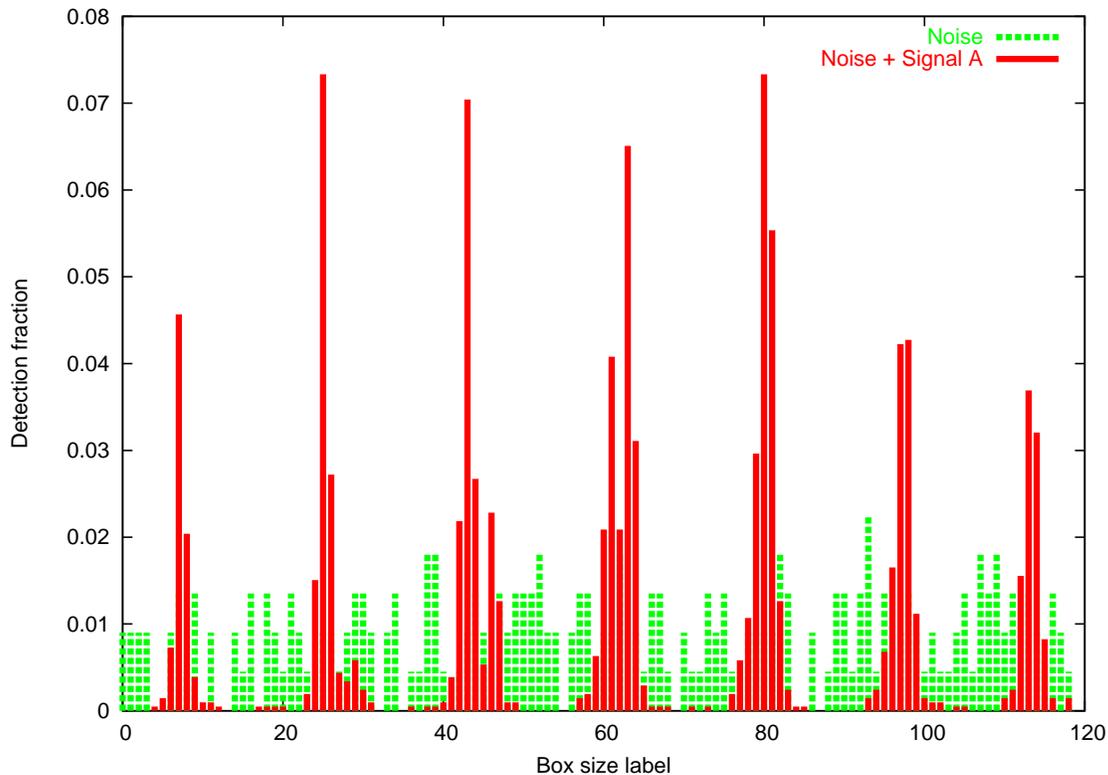}}
\caption{Fraction of detections in each box size when the signal A is present and absent. The box sizes have been numbered sequentially -- boxes numbered 1-17 have $n_t=1$, $n_f = 1,2,4...$, boxes numbered 18-34 have $n_t=2$, $n_f=1,2,4,...$ etc. Thresholds were assigned to give equal FAP$_j$ to each box size.}
\label{whichbin}
\end{figure}

If we are searching for a particular source, the threshold choice can be tuned. To illustrate this, we use the `typical' EMRI described in \cite{lisa5} (source ``A''). This is the last three years of the inspiral of a $10$ \msun\ black hole into a $10^6$ \msun\ SMBH of spin $a=0.8$M, with orbital eccentricity $e_0 = 0.4$ at the start of the observation and inclination $\iota_0 = 45^{{\rm o}}$. The frequency spread, time spread and SNR of our target waveform in principle allow us to guess theoretically which box size will be optimal for detecting it. However, the presence of noise means that the source will not always be detected in the same box size and indeed it will not always be the same part of the inspiral that is detected (different box sizes are optimal for different stages of the inspiral). To illustrate this, we ran a Monte Carlo simulation of the search over several hundred noise realizations with source A present in the data, and recorded every time a threshold was exceeded, and for which box size this occurred. The thresholds were set to give equal FAP$_j$ to all box sizes. When this procedure was repeated without the signal, the number of detections was roughly independent of the box size, as we would expect. In Figure~\ref{whichbin} we show the fraction of `detections' (i.e., a threshold being exceeded) as a function of the box size label. It is clear that there are preferred box sizes for detecting this source, but more than one. For a given noise realization, the source might be detected (i.e., exceed the box threshold $\rho_j$) with more than one box size, but these box sizes do vary from realization to realization. Thus, a targeted search for source A should still use a number of box sizes.

\begin{figure}
\centerline{\includegraphics[keepaspectratio=true,height=6in,angle=-90]{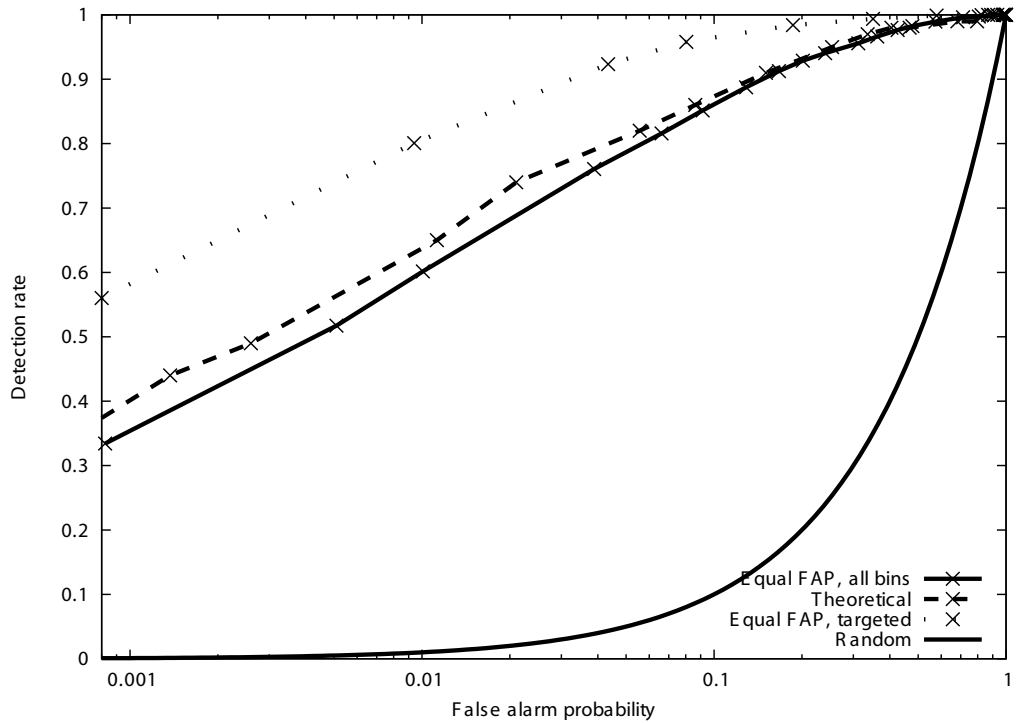}}
\caption{ROC of the search for detection of source A at $2$ Gpc under different assignments of the thresholds. ``Equal FAP, all bins'' is the threshold choice that assigns equal false alarm probability to all box sizes used. ``Theoretical'' is the threshold choice used in \cite{lisa5}, which assigns equal false alarm probability to all box sizes, assuming the maximum power is distributed as the maximum of $N/(m/4)$ independent points drawn from a $\chi^2_{4m}$ distribution. ``Equal FAP, targeted'' uses a reduced number of box sizes, tuned for this source, and assigns equal FAP$_j$ to each. ``Random'' indicates the performance of a random search, for which the detection rate and false alarm probability are equal.}
\label{compthresh}
\end{figure}

In Figure~\ref{compthresh} we compare the performance of the algorithm at detecting source A under these various choices for the thresholds. We show the Receiver Operator Characteristic (ROC) curve, which plots the detection rate versus the overall search FAP. The detection rate was estimated by adding the signal waveform to a sequence of noise realizations, and running the search algorithm. We used $\sim6500$ noise realizations, giving an accuracy in the detection rate of $\sim0.02\%
$.

The theoretical threshold assignment used in \cite{lisa5} gives performance that is comparable with the true `equal FAP$_j$' threshold assignment. A targeted search, which uses only the eleven box sizes which have detection rates greater than $0.03$ in Figure~\ref{whichbin}, and assigns equal FAP$_j$ to each, performs significantly better than the blind search. However, if we use the set of box sizes tuned for source A to detect other sources, the performance can be worse. For example, when searching for the source ``S'' (described below) at a distance of $1.2$ Gpc and at a search FAP of $1\%
$, the detection rate using thresholds tuned for source A is $25\%$ compared to $50\%
$ in the untargeted search. The search can also be tuned for source S, and a final data analysis scheme might involve using a targeted search for one type of inspiral followed by a targeted search for another type and so on. However, using multiple tuned searches increases the overall false alarm probability and eventually will do no better than a blind search. Without a much better knowledge of the likely parameters of EMRI sources in the LISA data, it is probably best to treat all box sizes equally. In the next section, we will use thresholds that give equal FAP$_j$ to all box sizes.

\section{Algorithm performance}
\label{algperf}
To assess the algorithm's performance, we computed ROC curves for a sequence of inspirals at a number of distances. The inspiral waveforms were computed using the `numerical kludge' approximation \cite{kostas02,tev05,gair05}, with detector modulations added using the low frequency approximation to the LISA response \cite{curt98}. The detector response can be treated more accurately \cite{cornish03,rubbo04,vallis05} but the low frequency response is sufficient for this analysis. Our `typical' source, ``A'', has the parameters specified above and was placed at a sky position $\cos(\theta_S)=0.5$, $\phi_S=1$, with orientation chosen as $\cos(\theta_K)=-0.5$, $\phi_K=4$. The angles $\theta_S$ and $\phi_S$ are the ecliptic colatitude and azimuth of the source on the sky, while the angles $\theta_K$ and $\phi_K$ are the ecliptic colatitude and azimuth to which the spin of the SMBH would point in the Solar System frame. The other trial waveforms were obtained by changing one or two of the inspiral parameters. The waveform parameters for the inspirals considered are listed in Table~\ref{waveparams}. For each inspiral, the initial semi-latus rectum, $p$, of the orbit was chosen such that the plunge would occur a few days after the end of the LISA observation, i.e., in each case we are focusing on approximately the last three years of inspiral. Table~\ref{waveparams} also lists the approximate signal to noise ratio that would be achieved in a fully coherent matched filtering search, using both Michelson data streams (I and II), if the source was at a distance of $1$ Gpc.

\begin{figure}
\centerline{\includegraphics[keepaspectratio=true,width=5in,angle=-90]{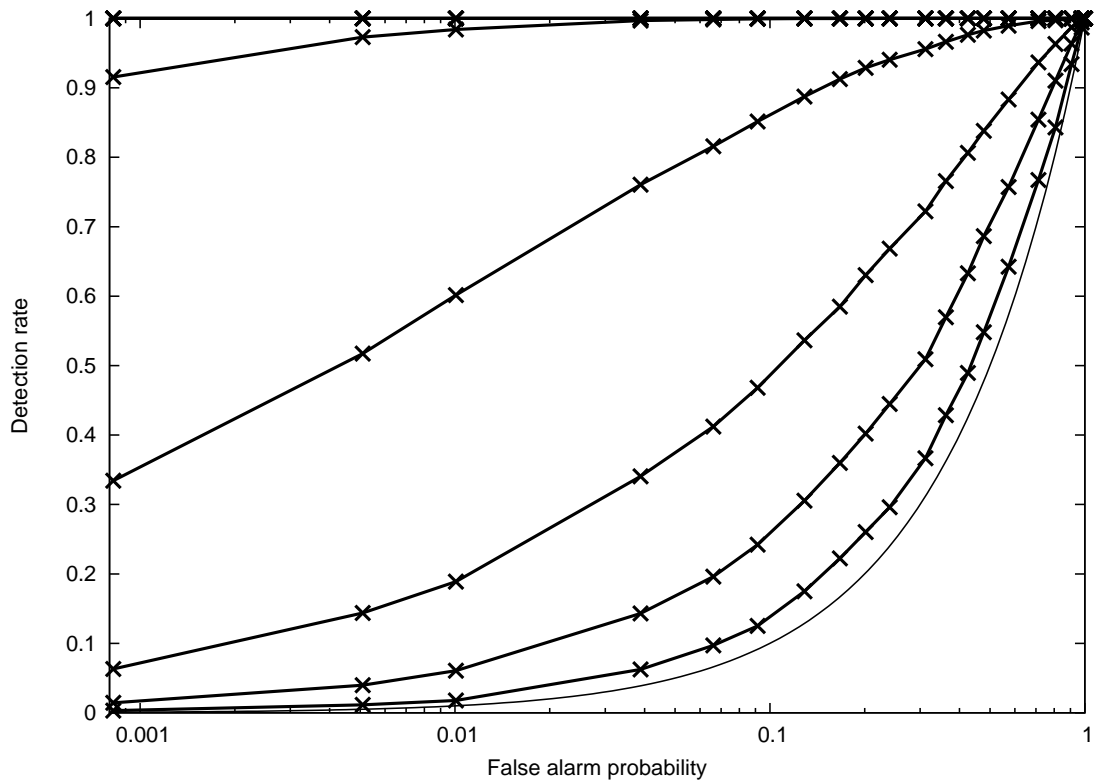}}
\caption{ROC performance of the search for detecting the typical EMRI, source ``A'', at distances of $0.8$, $1$, $1.2$, $1.4$, $1.75$, $2$, $2.25$, $2.5$ and $3$ Gpc (from uppermost downward). The lowermost solid line indicates the performance of a random search, as in Figure~\ref{compthresh}. Note that the curves for $0.8\--1.4$ Gpc lie on top of the upper axis, with $100\%$ detection rate at all the false alarm rates considered.}
\label{ROCSourceA}
\end{figure}

\begin{figure}
\centerline{\includegraphics[keepaspectratio=true,width=3in,angle=-90]{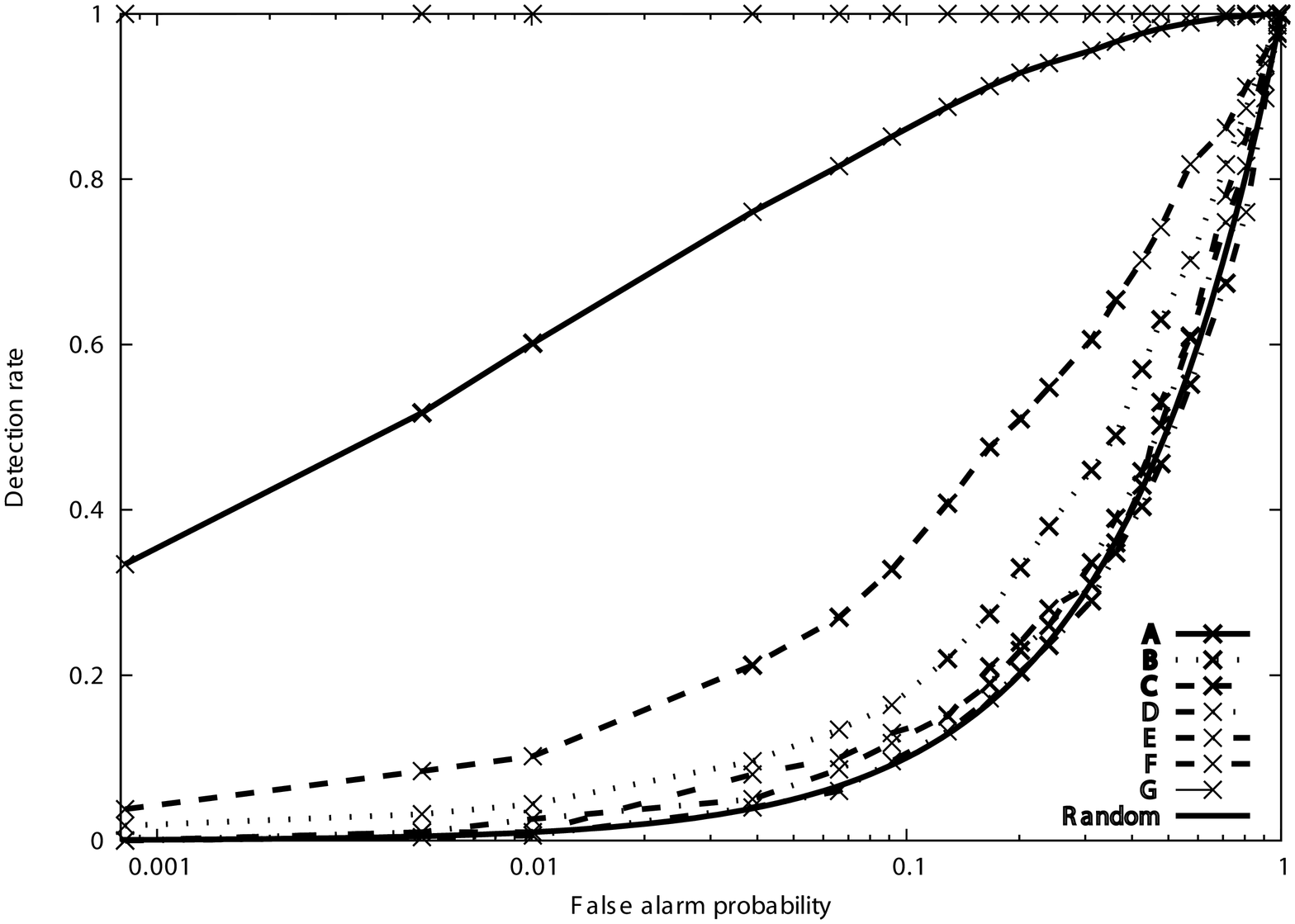} \includegraphics[keepaspectratio=true,width=3in,angle=-90]{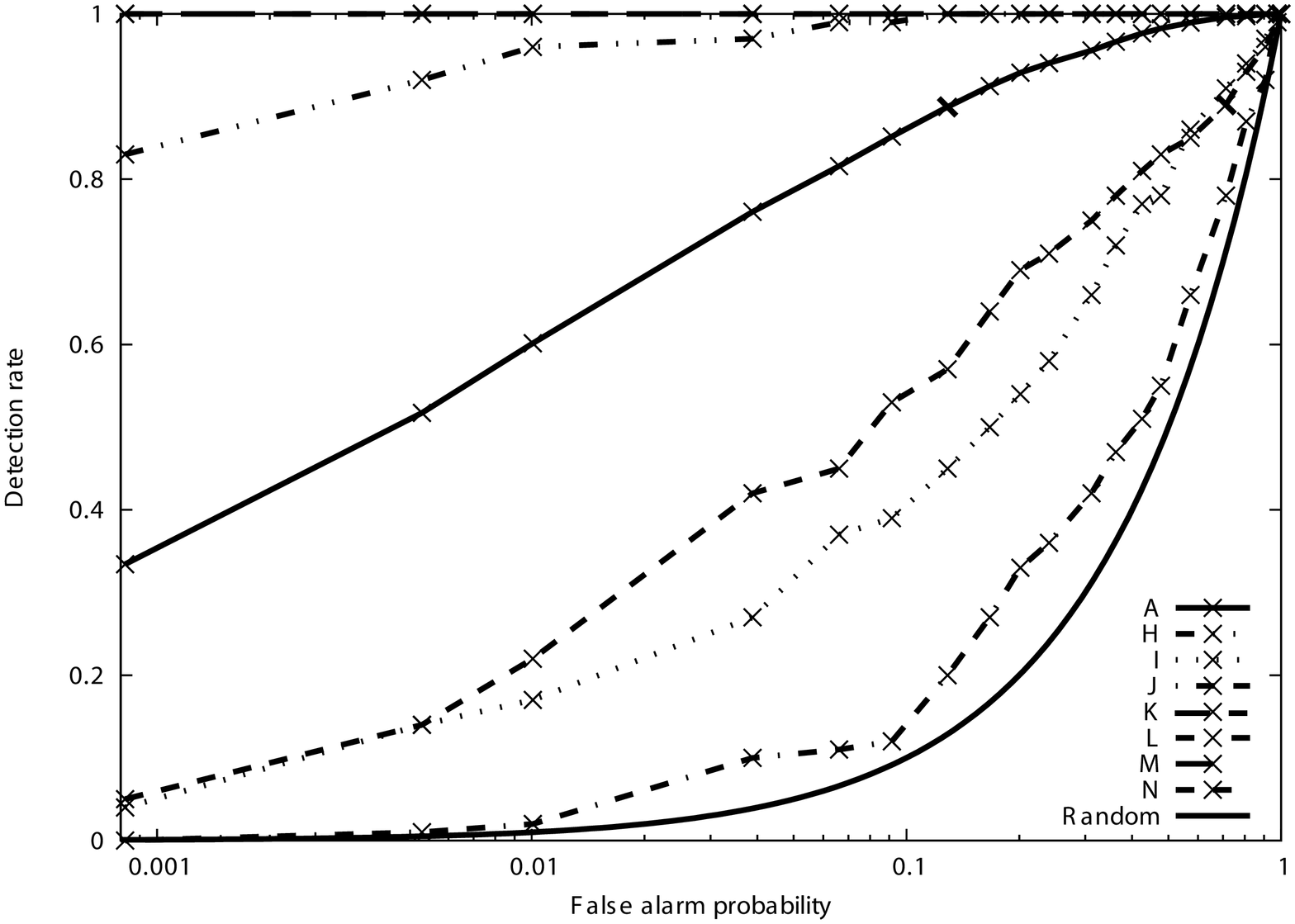}}
\centerline{\includegraphics[keepaspectratio=true,width=3in,angle=-90]{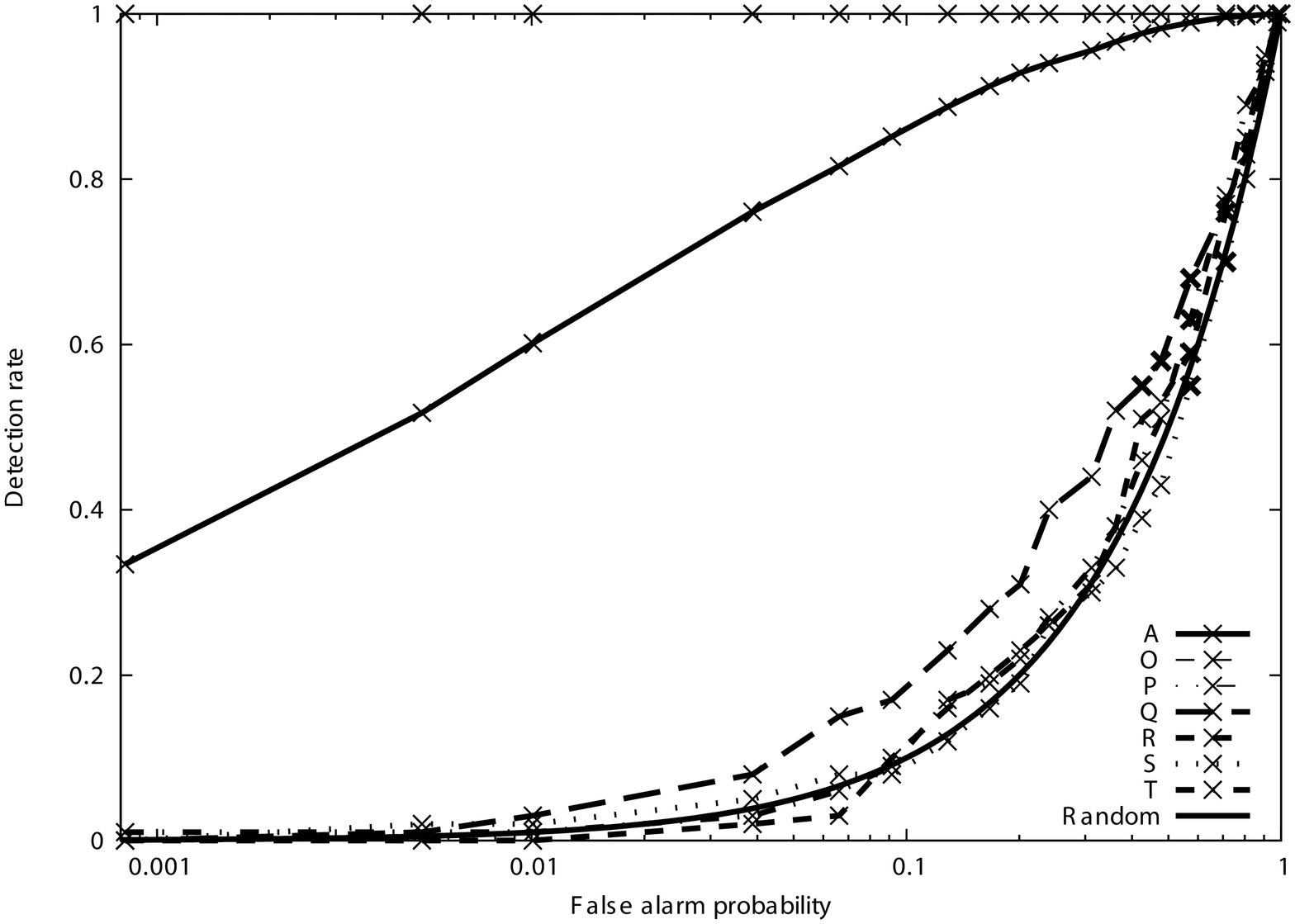} \includegraphics[keepaspectratio=true,width=3in,angle=-90]{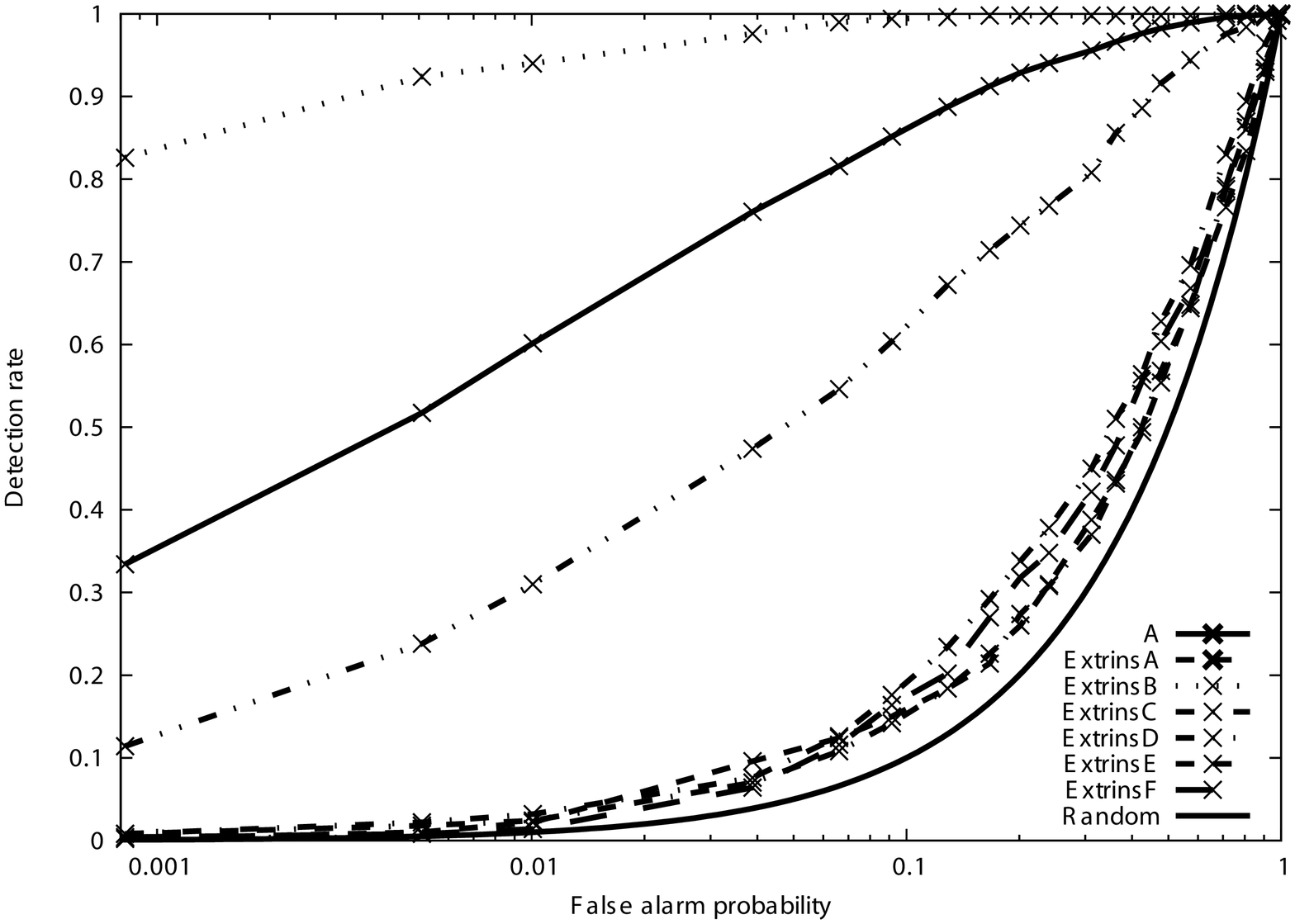}}
\caption{ROC performance of the search for detecting the trial EMRI events at a distance of $2$ Gpc. Each plot shows the typical inspiral source ``A'', and the ROC performance of a random search. The other trial sources have been divided among the four figures and have the parameters specified in Table~\ref{waveparams}.}
\label{ROCAllSource2Gpc}
\end{figure}

For each trial waveform we performed a Monte Carlo simulation, with the source at distances of $0.8$, $1.2$, $1.4$, $2$ and $3$ Gpc. Source ``A'' was additionally placed at distances of $1$, $1.75$, $2.25$ and $2.5$ Gpc. In Figure~\ref{ROCSourceA} we show the ROC for detection of source ``A'' at each distance, in Figure~\ref{ROCAllSource2Gpc} we show the ROCs for detection of all the sources at a distance of $2$ Gpc and in Table~\ref{detectrates}, we summarize the detection rates for each source at each distance, if the thresholds are set to give an overall search false alarm probability of $\sim1\%
$. Since this is the false alarm rate per LISA mission, we could afford to increase this number in the final search, up to $10\%
$ or more. However, for the present we use $1\%
$ in order to compare the performance of this technique directly to that of the semi-coherent matched filtering search \cite{jon04}. In the latter search, a matched filtering SNR of $\sim35$ is required for detection when the overall search FAP is set at $1\%
$. No ROC curves for the semi-coherent search are presently available, but the detection rate is estimated to be $\sim50\%
$ at threshold. 

With an overall search false alarm probability of $1\%
$, we see from Figure~\ref{ROCSourceA} that source A is easily detectable out to $2$ Gpc. The detection rate at greater distances decreases quite rapidly, although it is still $\sim20\%
$ at $2.25$ Gpc. The absolute limit of the search for source A is approximately $3$ Gpc, at which point we do no better than a random search. This compares quite favourably with the performance of the semi-coherent matched filtering search \cite{jon04}, which in principle can detect this source out to $\sim4.5$ Gpc in a Euclidean Universe. In fact, in \cite{jon04}, the reach of the search was limited at a redshift $z\approx1$, or $\sim3.5$ Gpc, by astrophysical considerations, since there is considerable uncertainty about the formation of EMRI sources at higher redshift.

Varying the parameters of the inspiral has a noticeable effect on the detectability. The white dwarf inspirals (sources D, E and F) are essentially not detectable (they would be detectable very close by, but we do not expect any such sources closer than $\sim0.5$ Gpc). However, this is because these sources are weak and the SNR is so low that they would not even be detected by the semi-coherent technique. The intermediate mass black hole inspiral (source G) can be detected to great distances since it is very loud, but it is presently unclear whether any such sources are likely to exist \cite{madaurees01,miller02}. The inspirals into a lower (source B) or higher (source C) mass SMBH are less detectable, the former becoming marginally detectable at $2$ Gpc and the latter becoming marginally detectable at about $2.25$ Gpc. This is a consequence of the shape of the LISA noise curve. Using the semi-coherent technique these sources are similarly less detectable than source A, although they can still be detected out to $\sim3.5$ Gpc. 

Varying the eccentricity (sources K--N) has a moderate effect on detectability -- lower eccentricity sources can be detected somewhat further away. This is in contrast to the semi-coherent technique, for which eccentricity has little effect on detectability. Low eccentricity inspirals have power spread over fewer frequency components, and so the power is more likely to be concentrated in certain regions in the t-f plane, making it easier for this algorithm to detect them. Increasing the central SMBH spin (sources H--J) appears to increase the detectability. Inspirals into highly spinning black holes will be at higher frequencies (and thus over the floor of the instrumental noise) and will plunge closer to the central SMBH. Both of these properties increase the SNR of the source and therefore are likely to increase the detectability by either this t-f method or the semi-coherent method, although this has not been investigated for the latter technique. Increasing the inclination of the orbit with respect to the equatorial plane of the SMBH (sources O--T) also affects detectability, with lower inclination orbits being more detectable. The inclination is defined \cite{kostas02} such that an inclination of $0^{{\rm o}}$ is a prograde equatorial orbit, while an inclination of $180^{\rm o}$ is a retrograde equatorial orbit. The increased detectability at low inclination is again largely due to the fact that such orbits plunge closer to the SMBH and thus have higher SNR. 

Finally, changing the extrinsic parameters (source position on the sky and inclination of the SMBH spin with respect to our line of sight) also affects detectability (sources ExtrinsA--ExtrinsF). By ensuring that the source is in a sky position to which LISA is very sensitive, and that the black hole spin is oriented favourably with respect to the detector, the detectability can be increased. Similarly, placing the source unfavourably can make detection marginal at $2$ Gpc. However, the extrinsic parameters chosen for source A seem to represent an `average' response performance.

\begin{table}
\begin{tabular}{|c|l|c|c|}
\hline Label&Parameters&Initial $p/M$&SNR \\ \hline \hline A&See text&$10.3$&$155$ \\ \hline B&$M=3\times10^5$\msun&$18.25$&$119$ \\ \hline C&$M=3\times10^6$\msun&$6.5$&$110$ \\ \hline D&$m=0.6$\msun, $M=3\times10^5$\msun&$9.405$&$14.1$ \\ \hline E&$m=0.6$\msun&$5.83$&$21.0$ \\ \hline F&$m=0.6$\msun, $M=3\times10^6$\msun&$4.511$&$15.7$ \\ \hline G&$m=100$\msun&$17.78$&$382$ \\ \hline H&$a=0.95M$&$10.07$&$170$ \\ \hline I&$a=0.5M$&$10.74$&$132$ \\ \hline J&$a=0.1M$&$11.31$&$108$ \\ \hline K&$e_0=0$&$10.42$&$147$ \\ \hline L&$e_0=0.1$&$10.41$&$150$ \\ \hline M&$e_0=0.25$&$10.385$&$151$ \\ \hline N&$e_0=0.7$&$9.71$&$159$ \\ \hline O&$\iota=0$&$9.925$&$223$ \\ \hline P&$\iota=30^{{\rm o}}$&$10.1$&$189$ \\ \hline Q&$\iota=60^{{\rm o}}$&$10.59$&$115$ \\ \hline R&$\iota=120^{{\rm o}}$&$12.126$&$57.7$ \\ \hline S&$\iota=150^{{\rm o}}$&$12.82$&$79.6$ \\ \hline T&$\iota=180^{{\rm o}}$&$13.11$&$87.0$ \\ \hline ExtrinsA&$\cos(\theta_S)=0.99$&$10.3$&$117$ \\ \hline ExtrinsB&$\cos(\theta_S)=0.01$&$10.3$&$162$ \\ \hline ExtrinsC&$\cos(\theta_K)=0.99$&$10.3$&$118$ \\ \hline ExtrinsD&$\cos(\theta_K)=0.01$&$10.3$&$146$ \\ \hline ExtrinsE&$\phi_K=0.01$&$10.3$&$111$ \\ \hline ExtrinsF&$\phi_K=2.$&$10.3$&$111$ \\ \hline
\end{tabular}
\caption{Parameters and signal to noise ratios at $1$ Gpc for trial waveforms. Unspecified parameters are the same as source ``A'', as given in the text.}
\label{waveparams}
\end{table}								     

\begin{table}
\begin{tabular}{|c||c|c|c|c|c|c|c|c|c|}
\hline &$0.8$ Gpc&$1.2$ Gpc&$1.4$ Gpc&$2$ Gpc&$3$ Gpc \\ \hline \hline A&$1$&$1$&$1$&$0.60$&$0.02$ \\ \hline B&$1$&$1$&$0.93$&$0.04$&$0.01$ \\ \hline C&$1$&$1$&$1$&$0.10$&$0.02$ \\ \hline D&$0.00$&$0.00$&$0.02$&$0.01$&$0.01$ \\ \hline E&$0.01$&$0.00$&$0.00$&$0.02$&$0.00$ \\ \hline F&$0.03$&$0.02$&$0.02$&$0.01$&$0.02$ \\ \hline G&$1$&$1$&$1$&$1$&$1$ \\ \hline H&$1$&$1$&$1$&$0.96$&$0.01$ \\ \hline I&$1$&$1$&$1$&$0.17$&$0.00$  \\ \hline J&$1$&$1$&$0.85$&$0.02$&$0.01$ \\ \hline K&$1$&$1$&$1$&$1$&$0.51$ \\ \hline L&$1$&$1$&$1$&$1$&$0.29$ \\ \hline M&$1$&$1$&$1$&$1$&$0.07$ \\ \hline N&$1$&$1$&$0.99$&$0.22$&$0.00$ \\ \hline O&$1$&$1$&$1$&$1$&$0.63$ \\ \hline P&$1$&$1$&$1$&$1$&$0.10$ \\ \hline Q&$1$&$1$&$0.85$&$0.03$&$0.00$ \\ \hline R&$0.8$&$0.02$&$0.04$&$0.00$&$0.01$ \\ \hline S&$1$&$0.53$&$0.1$&$0.02$&$0.02$ \\ \hline T&$1$&$0.96$&$0.36$&$0.01$&$0.01$ \\ \hline ExtrinsA&$1$&$1$&$0.65$&$0.02$&$0.01$ \\ \hline ExtrinsB&$1$&$1$&$1$&$0.94$&$0.03$ \\ \hline ExtrinsC&$1$&$1$&$0.82$&$0.03$&$0.02$ \\ \hline ExtrinsD&$1$&$1$&$1$&$0.31$&$0.02$ \\ \hline ExtrinsE&$1$&$0.99$&$0.57$&$0.03$&$0.02$ \\ \hline ExtrinsF&$1$&$0.99$&$0.62$&$0.02$&$0.02$ \\ \hline
\end{tabular}
\caption{Detection rates for trial waveforms at various distances. Thresholds were set using the numerical probability distributions, and with an overall search false alarm probability of $1\%$.}
\label{detectrates}
\end{table}

\section{Conclusion}
\label{conc}
We have presented a thorough analysis of the time-frequency method to detect EMRIs with LISA that was originally proposed in \cite{lisa5}. We have accurately computed the statistics of this search and examined its performance on a wide range of possible EMRI signals. We find that this algorithm is able to detect many different EMRI events out to distances of $1\--3$ Gpc, depending on the source parameters. In an untargeted search, a typical source can be detected at $2$ Gpc with a detection rate of $60\%
$ at a search false alarm probability of $1\%
$. Lower eccentricity sources, which have less frequency spreading, can be detected as far away as $3$ Gpc with detection rates of $50\%
$ at the same overall FAP. The reach of the search can be extended by increasing the allowed FAP or by using a targeted search. By comparison, the semi-coherent matched filtering algorithm \cite{jon04} can reach $\sim4.5$ Gpc for an overall FAP of $1\%
$, but at a presently undetermined detection rate (perhaps $\sim50\%
$). Broadly speaking, this time frequency search has better than half the reach of the semi-coherent search \cite{jon04}, but at a tiny fraction of the computational cost. This is very encouraging, given the simplicity of the technique, and suggests that a method like this could be a valuable first step for detecting the loudest EMRI events in the LISA data. The approach is very similar to the `excess power' technique used in LIGO \cite{anderson01}, which was designed to search for bursting signals. Although we have concentrated on EMRIs, this technique will also be sensitive to other types of source, including bursts, galactic white dwarf binaries etc. and so it may have many applications.

However, one significant issue that we have not yet considered is that of confusion. The LISA data will be dominated by a superposition of signals from many different astrophysical sources. So far, we have only examined the problem of detecting a single source in instrumental noise (plus an astrophysical foreground of white dwarf binaries). This algorithm should be able to make detections when there are multiple sources if the sources are reasonably well separated in time and frequency or of widely different signal to noise ratios, but it is not clear how it will fair on overlapping sources of comparable brightness. It is likely that the performance will be significantly impaired in such cases, due to the simplicity of the algorithm. Additionally, a shortcoming of this simple technique is that while it can tell us that a source is present, it does not provide much information about the properties or parameters of that source. In principle, a time-frequency analysis can provide at least the frequency and rate of frequency drift of an inspiral source, but this requires more sophisticated algorithms. One possibility would be to use boxes of different (i.e., non-rectangular) shapes, or to use a pattern recognition algorithm in the (binned or unbinned) t-f plane to search for tracks corresponding to inspirals. These more sophisticated techniques should also cope more readily with confusion. Finally, it will be instructive to test this and future algorithms using more accurate inspiral waveforms. The simple kludged waveforms used here are purely quadrupolar, and so their frequency structure is not as complex as true EMRI waveforms which might influence our results.

Despite these remaining issues, the performance of this algorithm is very promising and suggests that a time-frequency method of some kind is likely to be a useful first step in detecting EMRIs and other types of source in the LISA data. This provides a good incentive for developing and testing t-f algorithms in the ways suggested above and this work is now underway.

\ack We thank Curt Cutler, Teviet Creighton, Leor Barack and Kip Thorne for critical discussions of this work. This work was supported in part by the Max Planck Institut fuer Gravitationshphysik (Albert Einstein Institut) (LW), by NASA grants NAG5-12834 and NAG5-10707 (JG) and by St. Catharine's College, Cambridge (JG).


\section*{References}
\begin{thebibliography}{10}
\bibitem{lisa5} Wen L and Gair J R 2005 \CQG {\bf 22} S445
\bibitem{jon04} Gair J R, Barack L, Creighton T, Cutler C, Larson S L, Phinney E S and Vallisneri M 2004 \CQG {\bf 21} S1595
\bibitem{leor04} Barack L and Cutler C 2004 \PR D {\bf 69} 082005 
\bibitem{nelemans01} Nelemans G, Yungelson L R and Portegies Zwart S F 2001 {\it Astronomy \& Astrophysics} {\bf 375} 890
\bibitem{timpano05} Timpano S E, Rubbo L J and Cornish N J 2005 {\it preprint} gr-qc/0504071
\bibitem{curt98} Cutler C 1998 \PR D {\bf 57} 7089 
\bibitem{kostas02} Glampedakis K, Hughes S A and Kennefick D 2002 \PR D {\bf 66} 064005
\bibitem{tev05} Creighton T, Gair J R, Hughes S A and Vallisneri M 2005, in preparation
\bibitem{gair05} Gair J R and Glampedakis K 2005, in preparation
\bibitem{cornish03} Cornish N J and Rubbo L J 2003 \PR D {\bf 67} 022001, Erratum-ibid. D {\bf 67} 029905
\bibitem{rubbo04} Rubbo L J, Cornish N J and Poujade O 2004 \PR D {\bf 69} 082003
\bibitem{vallis05} Vallisneri M 2005 \PR D {\bf 71} 022001
\bibitem{madaurees01} Madau P and Rees M J 2001 {\it Astrophys. J. Lett.} {\bf 551} L27
\bibitem{miller02} Miller M C 2002 {\it Astrophys. J.} {\bf 581} 438
\bibitem{anderson01} Anderson W G, Brady P R, Creighton J D E and Flanagan E E 2001 \PR D {\bf 63} 042003
\end{thebibliography}
\end{document}